**Imaging and characterization of spontaneous vortices in a proximity-induced superconductor**

Iku Nakaaki,[1†] Kotaro Taki,[1] Mio Nomura,[1] Haruna Ishimaru,[1] Minagi Yono,[1] Jun Chen,[2] Hiroyo Segawa,[2] Akiko Nakamura,[3] Taku Moronaga,[3] Minoru Tachiki,[4] Shuuichi Ooi,[4] Shunichi Arisawa,[5] Tsutomu Nojima,[6] and Takashi Uchino[1*]

[1]*Department of Chemistry, Graduate School of Science, Kobe University, Nada, Kobe 657-8501, Japan*
[2]*Research Center for Electronic and Optical Materials, National Institute for Materials Science, Tsukuba, Ibaraki 305-0044, Japan*
[3]*Research Network and Facility Services Division, National Institute for Materials Science, Tsukuba, Ibaraki 305-0047, Japan*
[4]*International Center for Materials Nanoarchitectonics (MANA), National Institute for Materials Science, Tsukuba, Ibaraki 305-0047, Japan*
[5]*Research Center for Functional Materials, National Institute for Materials Science, Tsukuba, Ibaraki 305-0047, Japan*
[6]*Institute for Materials Research, Tohoku University, Sendai, Miyagi 980-8577, Japan*

Observation of spontaneous symmetry breaking is crucial for understanding continuous second-order phase transitions from disordered to ordered states, which often leads to the formation of topological defects. In superconductors, such topological defects manifest as quantized vortices. However, the formation and observation of spontaneous vortices in a uniform superconductor are challenging because extremely rapid cooling (>$10^8$ K/s) is generally required for that purpose. Here we conducted scanning superconducting quantum interference device microscope (SSM) measurements on an $MgB_2$-based proximity-induced superconductor, an intrinsically inhomogeneous system. In this system, individual superconducting domains will reach internal equilibrium independently during cooling and choose their own phase before the global phase coherence is established via the long-range proximity coupling. The SSM measurements demonstrate that vortices are nucleated spontaneously even at a relatively slow cooling rate (~0.2 K/s). We also find that the vortices with different polarities, sizes, and shapes appear stochastically under near-zero-field conditions. The geometry of the spontaneous vortices is more extended than that of the field-induced Abrikosov vortices. Magnetic field profile analysis based on the London model elucidates that penetration depths of the extended vortices are anomalously large, exceeding several micrometers. This unusual morphology of the spontaneous vortices most likely imprints the information that is frozen at the moment of vortex formation. Our findings not only provide insights into the local phase differences present in the early stage of the phase transition in this proximity-induced superconducting system, but they also shed insights into the structure, formation, and stabilization of topological defects in highly disordered and inhomogeneous superconducting systems.

## I. INTRODUCTION.

The spontaneous formation of topological defects is a universal phenomenon that arises when a system is driven into a phase of broken symmetry. This topic has been explored in diverse fields, ranging from the formation of cosmic strings [1–4] to spontaneous vortex formation in Bose-Einstein condensates [5–9], superconductors [10–13], and quantum transport systems [14–16]. The Kibble-Zurek (KZ) mechanism [1,17,18] is a paradigmatic scenario for describing the formation of topological defects (or vortices) during cooling. The KZ scenario assumes that different regions with different phases cannot communicate rapidly enough to maintain equilibrium, which leads to the defect formation based on the causal mechanism. Consequently, the phase information is expected to be frozen into the defects. The phase $\phi$ is correlated over a characteristic length $2\pi\bar{\xi}$ where $\bar{\xi}$ scales with the quench time $\tau_Q$ as $\bar{\xi} \propto \tau_Q^{\alpha}$ , with $\alpha$ being the KZ scaling exponent [18]. Hence, the number density and spatial distribution of defects will carry information about the dynamics of the phase transition [19]. Accordingly, most previous studies have focused on the dependence of defect density and correlation statistics on quench time, as demonstrated in superconducting rings [10], a multi-Josephson-junction loop [20], annular Josephson tunnel junctions [12,21,22] and thin films [11,13]. These results, together with observations in other systems [8,9,23–25], have stimulated extensive efforts to generalize the KZ mechanism, leading to substantial theoretical developments [26–29]. In addition to the above approaches, it could be possible to infer the length scale of a frozen phase by analyzing the shape and size of the resulting vortices using appropriate magnetic imaging techniques [30,31]. Thus far, however, less attention has been paid to the geometric features of the spontaneous vortices in superconductors because a geometrical analysis based on their direct imaging is experimentally challenging.

† Present address: AGC Ceramics Co., Ltd, Takasago, Hyogo 676-8655, Japan
* Contact author: uchino@kobe-u.ac.jp

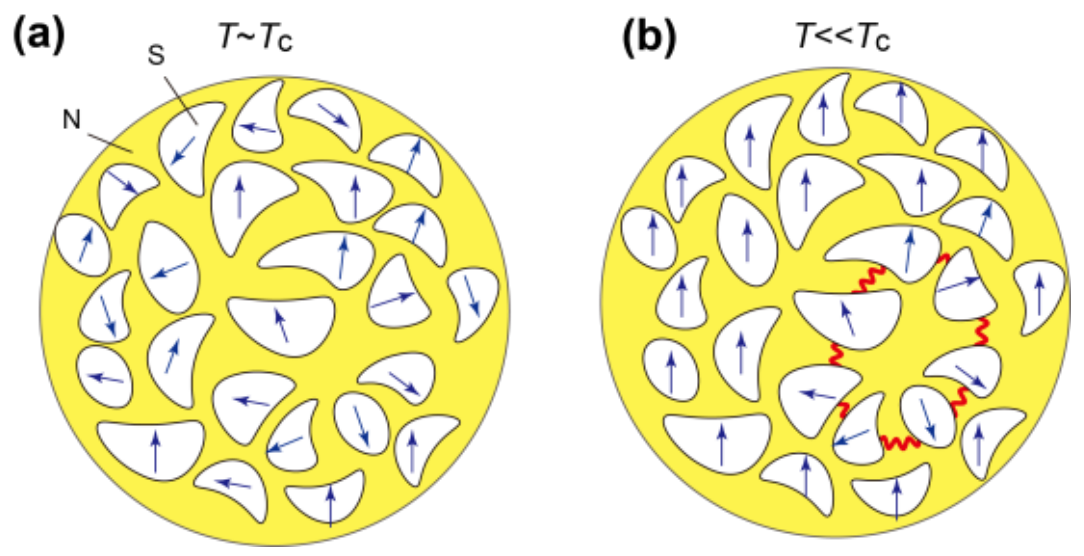

FIG. 1. Schematic representation of a proximity-induced superconducting nanocomposite. (a) At temperatures near the transition temperature $T_c$, each superconducting (S) domain embedded in the normal (N) matrix reaches internal equilibrium with a different phase (represented by arrows), meaning that the respective S domains are uncoupled. (b) At temperatures well below $T_c$, the entire system goes towards a global phase coherence state, in which the respective S domains have an identical phase state, owing to the proximity effect, although topological defects where the sum of the phases around a loop (represented by a red wavy circle) exceeds $\pm 2\pi$ may be created during a thermal quench.

Here we image spontaneous vortices nucleated in a proximity-induced superconducting composite system [for schematic representation, see Fig. 1(a)] using scanning superconducting quantum interference device (SQUID) microscopy (SSM). Because our composite consists of randomly distributed superconducting segments coupled via the proximity effect, the composite can be viewed as a three-dimensional (3D) ensemble of proximity-induced Josephson junctions [32,33]. This system will provide an experimental platform for studying the formation and stabilization of topological defects in highly disordered and inhomogeneous superconductors. Our scenario for the spontaneous vortex formation is as follows. Just below the transition temperature $T_c$ of the superconducting phase, each superconducting (S) domain embedded in the normal (N) matrix reaches internal equilibrium with an independent phase [Fig. 1(a)]. As the temperature decreases further below $T_c$, the critical currents of the proximity junctions increase. Eventually, the global phase coherence is established as a result of the formation of superconducting loops of different area and orientations [34,35]. If there exists a loop in which the sum of the phase differences exceeds $\pm 2\pi$, vortices can be nucleated spontaneously [Fig. 1(b)]. From the SSM observation on the proximity-coupled superconducting nanocomposite, we found that the vortices are nucleated stochastically under near-zero-field conditions even at a relatively slow cooling rate (~0.2 K/s), consistent with this scenario. The resulting vortices are highly extended and intricate in shape, in stark contrast to the field-induced Abrikosov vortices. The quantitative analysis of the field profile based on the London model revealed that the resulting vortices have anomalously large penetration depths of up to several micrometers. The observed geometrical features most likely convey the phase inhomogeneity that becomes frozen at the moment of vortex formation in this proximity-coupled system. We further explore the field distributions of these geometrically extended vortices using the London model.

## II. EXPERIMENTAL PROCEDURES

In our experiment, we first prepared a proximity coupled superconducting nanocomposite consisting of $MgB_2$ (~30 vol. %) and MgO (~70 vol. %) nanocrystals [Fig. 2(a)] by the solid-phase reaction of Mg and $B_2O_3$ powders under Ar atmosphere at 700°C, followed by a subsequent spark plasma sintering (SPS) procedure, according to the procedures reported previously [36–38] (for details, see Methods in Supplemental Material [39]). As has been demonstrated in our recent works [36–38], this MgO/$MgB_2$ nanocomposite exhibits bulk-like superconducting properties, including zero resistivity [Fig. 2(b)], perfect diamagnetism [Fig. 2(c)], and large critical current densities [Fig. 2(d)], irrespective of the low volume fraction (~30 vol. %) of $MgB_2$. From the temperature dependence of the real $4\pi\chi'$ and imaginary $4\pi\chi''$ parts of the ac susceptibility [Fig. 2(c)], one sees that the intergranular phase coherence begins to emerge at temperature $T$~38 K, and nearly perfect diamagnetism ($4\pi\chi' \sim -1$) is achieved for $T$ <~36 K. Note also that the current-voltage ($I$-$V$) curves in the temperature region near $T_c$ resemble those of an overdamped Josephson junction with thermal fluctuation [48,49] [Fig. 2(e)]. In our previous papers [37,38], we have demonstrated that the lower ($H_{c1}$) and the higher ($H_{c2}$) critical fields of this MgO/$MgB_2$ nanocomposite are 580 Oe and 100 kOe, respectively. The resulting zero-temperature Ginzburg-Landau (GL) coherence length $\xi(0)$ and penetration depth $\lambda(0)$ are estimated to be ~6 nm and ~90 nm, respectively [37,38], which are almost comparable to those of pure $MgB_2$ polycrystals [50]. These findings, although counterintuitive, implies that at sufficiently low temperatures, the strong and global phase coherence is achieved in this nanocomposite despite the high volume fraction (~70 vo. %) of MgO.

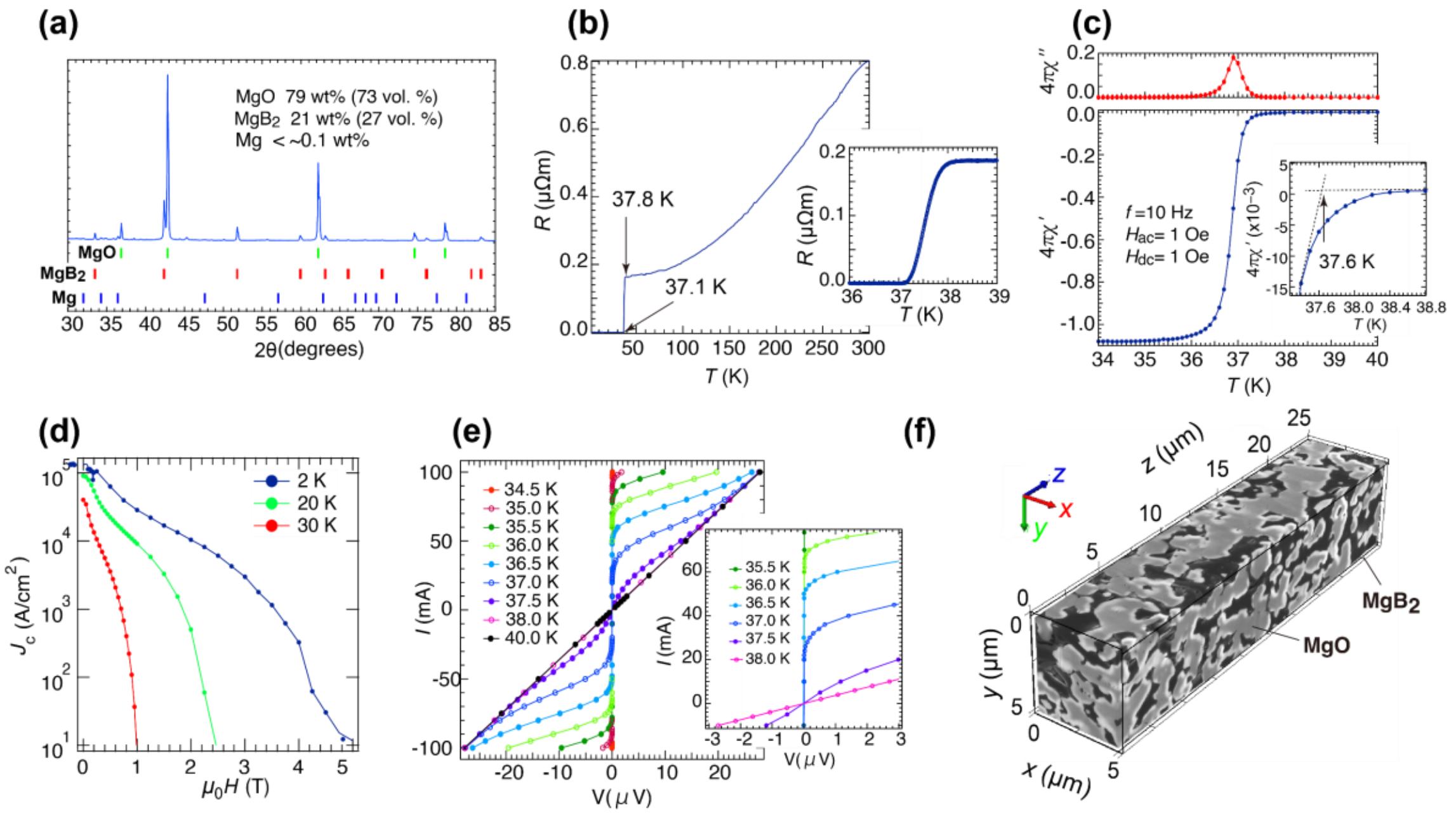

FIG. 2. Structural and superconducting properties of the superconducting nanocomposite. (a) XRD pattern and the weight fractions of MgO, $MgB_2$ and Mg obtained from a Rietveld analysis along with the corresponding approximate volume fractions. (b) Temperature-dependent resistivity. The inset shows an expanded plot of resistivity around the transition temperature $T_c$. (c) Temperature dependence of the real ($4\pi\chi'$) and imaginary ($4\pi\chi''$) parts of the ac susceptibility observed in a dc magnetic field $h_{dc}$ of 1 Oe using an ac frequency $f$ of 10 Hz at an ac amplitude $h_{ac}$ of 1 Oe. The inset shows an expanded plot of $4\pi\chi'$ around $T_c$. (d) Critical current density measured at 2, 20 and 30 K obtained from the magnetic hysteresis loops. (e) The current($I$)-voltage ($V$) curves measured in the temperature range of 34.5-40 K. The inset shows an expansion of the plot near the origin. (f) 3D image reconstructed from sequential focused ion beam scanning electron microscopy (FIB-SEM) images (For 3D animation, see Movie S1 in [39]). The black and gray regions correspond to the $MgB_2$- and MgO-rich regions, respectively, and a typical box-counting fractal dimension obtained from binary-converted SEM images is 1.68 (See also Fig. S1 in [39]).

The following three structural and interfacial properties specific to this nanocomposite are likely to be responsible for the observed hierarchical and robust long-range proximity effect (for more details, please refer to our previous publications [36–38]): (i) scale-free (or fractal) distribution of $MgB_2$ components [Fig. 2(f), and Fig. S1 and movie S1 in Supplemental Material [39]), (ii) metallic-like conductivity of the MgO phase due to a large number of oxygen vacancies introduced during the sample preparation process (Fig. S2 in Supplemental Material [39]), and (iii) the atomically clean $MgO/MgB_2$ interfaces (Fig. S3 in Supplemental Material [39]).

The thus prepared sample was cut into a plate shape with dimensions of 5 mm×5 mm×1 mm and then mirror-polished by diamond paste for subsequent SSM measurements. The surface roughness of the polished sample was found to be below 1 μm (see Fig. S4 in Supplemental Material [39]). This implies that there exists an uncertainty $\Delta z$ in the height of the SQUID pickup loop to the sample surface ($\Delta z \pm 0.5\,\mu m$). However, this uncertainty is small enough not to change the outcome of this work.

SSM images were obtained by using a scanning SQUID microscope (SQM-2000SII Nanotechnology) equipped with a conventional planar SQUID pickup loop with a diameter of 10 μm. A residual background magnetic field $H$ was $H$~15.0 mOe, which was compensated by the magnetic field applied from a Helmholtz coil to yield a near zero magnetic field of less than ~0.25 mOe. All the field values shown below are given after compensating for this background field. In the present SSM system, the cooling rate $R$ in the vicinity of $T_c$ was $R$~0.2 K/s, which is relatively slow as compared with those employed in previous experiments ( $R = 10 - 10^8$ K/s) [11, 12, 22–24]. Details of the measurements are given in Methods in Supplemental Material [39].

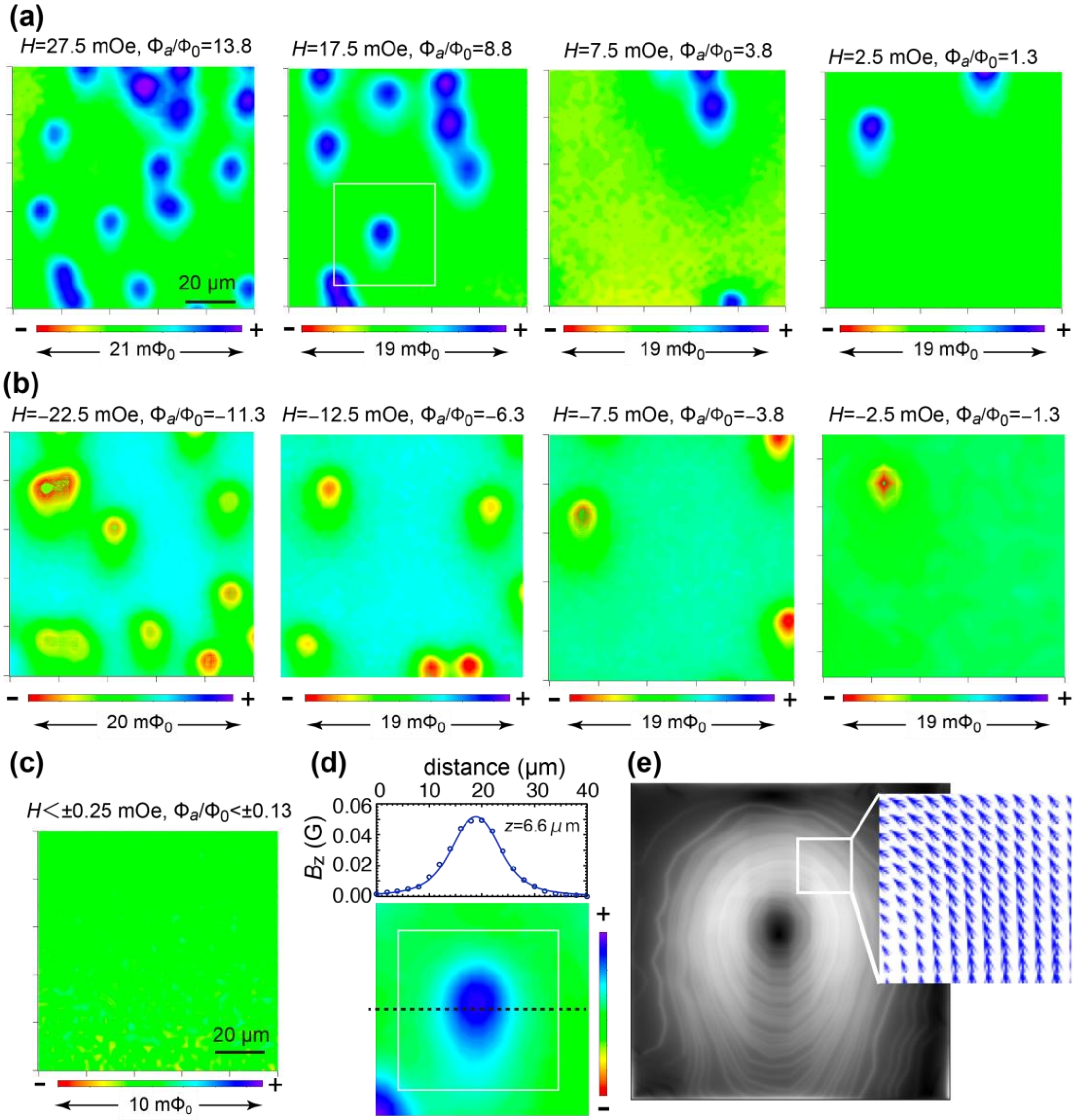

FIG. 3. Scanning SQUID microscope (SSM) images of the superconducting nanocomposite taken at a temperature of 4 K over the same area of 100 × 100 μm$^2$ under (a) positive field-cooling FC, (b) negative FC, and (c) nearly zero applied field ($H$≤±0.25 mOe) conditions with a pixel size of 2 × 2 μm$^2$. (d) The enlarged plot of the white box region in (a) (lower panel) and the field profile measured along the broken line in the SSM image (upper panel). In the upper panel, the solid line is a fitting based on the monopole model. (e) Local current distribution of the white box region in (d). The inset shows the calculated current of the white box region in the form of current vectors.

## III. RESULTS

### A. Scanning SQUID microscope imaging of vortices

Figures 3(a)–3(c) show a series of SSM images taken at $T = 4\ K$ over the same area of 100 × 100 μm$^2$ under positive field-cooling (FC) [Fig. 3(a)], negative FC [Fig. 3(b)], and nearly zero applied field [$H$≤±0.25 mOe, Fig. 3(c)] conditions. The number of vortices is in reasonable agreement with the applied flux $\Phi_a$ in units of a magnetic flux quantum $\Phi_0$ , i.e., $\frac{\Phi_a}{\Phi_0}$ , implying that each vortex has a single flux quantum. In Figs. 3(a) and 3(b), one sees round but slightly tear-

drop-shaped vortices with a lateral size of ~10 μm. The observed tear-drop shapes result from a convolution effect between the magnetic field lines of the vortex and the shape of the pickup loop [51.52]. Note also that, in the present SSM system, the pickup loop is positioned at a height *z* of several micrometers above the sample surface (*z*=6–7 μm, as will be shown below) and hence measures the spatial distribution of the *z*-component of the expanded magnetic field. The resulting spatial resolution of this SSM system is ~2.5 μm [47].

Since the penetration depth λ of this nanocomposite at 4 K (λ < ~100 nm) [37,38]) is expected to be much smaller than *z*, we can estimate the value of *z* by analyzing the field profile of the vortices on the basis of a monopole model, which is valid in the limit $\lambda \ll \sqrt{r^2+z^2}$ [51,53–56] (see Appendix A),

$$B_z(r,z) = \frac{\Phi_0}{2\pi}\frac{z+\lambda}{[r^2+(z+\lambda)^2\}^{\frac{3}{2}}}, \quad (1)$$

where $B_z$ is the magnetic field perpendicular to the surface of the sample and *r* is the horizontal distance from the vortex center. Provided that $\lambda \ll z$ in Eq. (1), the *z* value was obtained by fitting the field profile of the vortices to Eq. (1). An example of the field profile of an isolated vortex together with the result of the monopole fit is shown in Fig. 3(d). During fitting, we took into account the effect of the shape and area of the SQUID pickup loop on the observed $B_z(r,z)$ value (see also Appendix A). For comparison, the *z* values were also obtained by using a more accurate London model (details of the London model will be given below). We confirmed that both the monopole and London models yield practically the same fitted values of *z*. In the present experimental setup, the *z* value was finally determined to be 6.6±0.3 μm by averaging the fitted values of *z* obtained for six independent field-induced vortices (see Fig. S5 in Supplemental Material [39]).

Although the actual length scale of the vortices cannot be directly estimated from the present SSM images, it is interesting to visualize the current distribution of an isolated vortex using a numerical inversion scheme of Biot-Savart law (for details, see Methods in Supplemental Material [39]). An example of the vortex current is given in Fig. 3(e). One can recognize typical features of a vortex with positive polarity, namely, a central current-free core region and its surrounding circulating current in the anti-clockwise direction.

### B. Vortex formation under near-zero fields

As far as the limited area of 100 × 100 μm² shown in Fig. (3) is concerned, the SSM image taken under near-zero field showed no contrast (i.e., no vortices), as shown in Fig. 3(c). Even under such near-zero field conditions, however, the total applied flux $\frac{\Phi_a}{\Phi_0}$ over the whole scanned area may exceed unity when we examine a sufficiently large scanned area. Figures 4(a) and 4(b) shows the SSM images taken at *T* = 4 K over the area of 300 × 300 μm² under positive FC [*H*=5 mOe, $\frac{\Phi_a}{\Phi_0}$ =22.5; Fig. 4(a)] and near-zero field [*H*≤±0.25 mOe, $\frac{\Phi_a}{\Phi_0}$≤±1.1; Fig. 4(b)] conditions. In Fig. 4(a), one sees a few tens of vortices, as expected from the applied flux of $\frac{\Phi_a}{\Phi_0}$=22.5. As for the SSM image taken near-zero field conditions [Fig. 4(b)], one can also recognize vortices, whose number (~10) exceeds the upper bound of the estimated applied flux ($\frac{\Phi_a}{\Phi_0}$≤1.1). This suggests that not all of the vortices shown in Fig. 4(b) are driven by the external magnetic field, but some of them are generated without an applied field. This assumption is supported by the fact that in Fig. 4(b), the vortices with positive and negative polarities coexist within the same SSM image. If the vortices are induced by the applied field, only the vortices with the same polarity as that of the applied field are observed. It should also be noted that the vortices in Fig. 4(b) are appreciably more extended than those shown in Fig. 4(a), which can be more easily recognized by comparing the enlarged images shown in the lower panels of Figs. 4(a) and 4(b). It is important to note that the SSM images shown in Figs. 4(a) and 4(b) were acquired over exactly the same region of the sample with the identical condition (the sensor height *z* and temperature *T*) of the SSM system except for *H*. Therefore, the pronounced difference in the apparent vortex size – especially as seen in the enlarged images in the lower panels – cannot be attributed to the local variation of *z* or any surface-related phenomena. This strongly indicates that the extended vortices observed under near-zero-field conditions are not due to any artifact arising from uncontrolled experimental factors. Note, however, that upon closer inspection of Fig. 4(a), some of the vortices, such as those indicated by red arrows, appear to be larger than the others. This implies that although rare, spatially extended vortices can be formed even under applied fields, as will be discussed later.

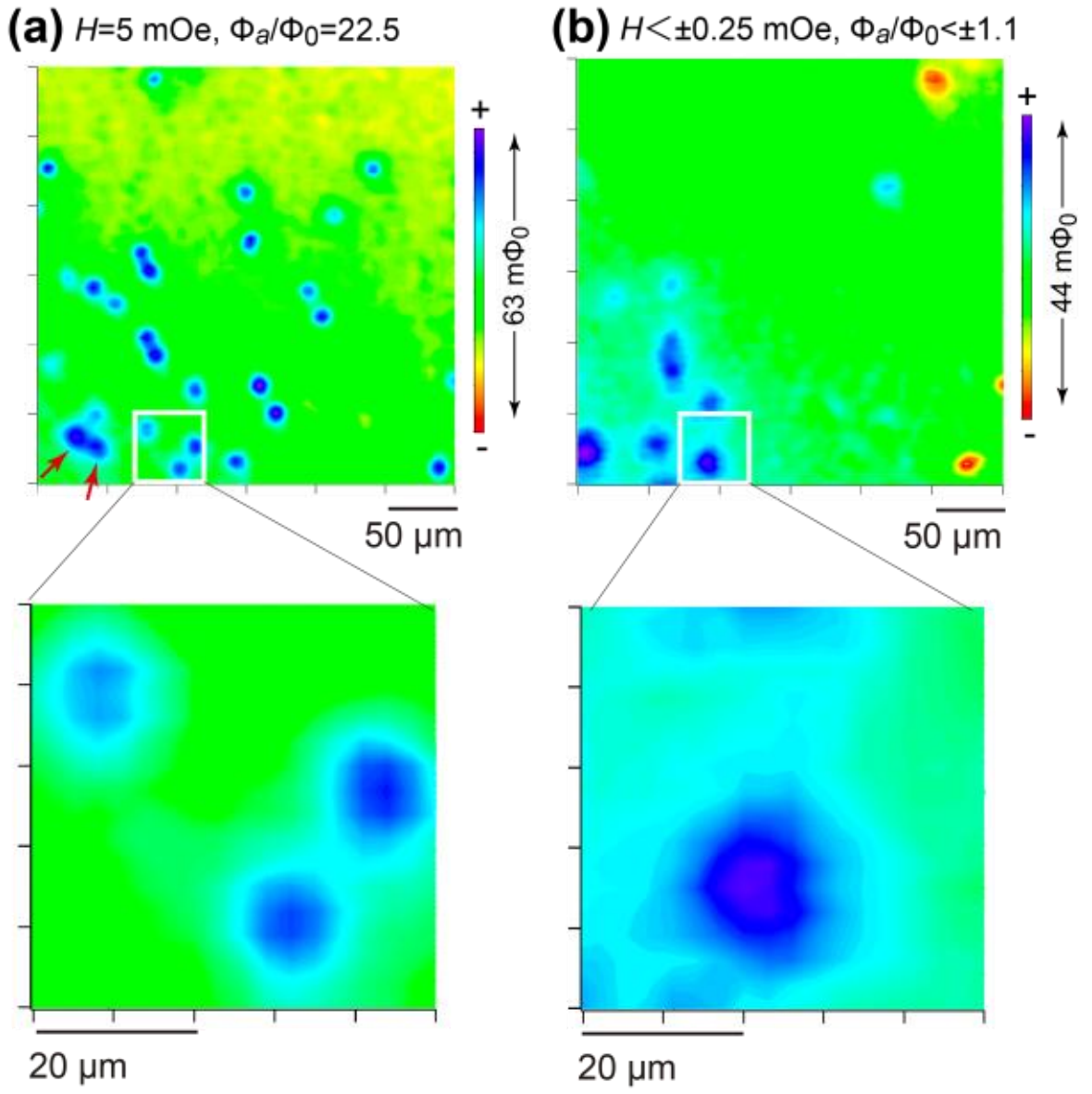

FIG. 4. SSM images taken at a temperature of 4 K over the same area of 300 × 300 μm² under (a) positive FC (*H*=5 mOe), and (b) nearly zero applied field (*H*≤±0.25 mOe) conditions with a pixel size of 3 × 3 μm². The lower panels show the enlarged images of the corresponding white box regions (50 × 50 μm²). The vortices indicated by the red arrows in (a) are the vortices that are more extended in shape than those found commonly in (a).

After a number of near-zero field SSM observations at 4 K over different areas, we acquired the SSM image showing the vortex images of both polarities simultaneously, as shown in Fig. 5(a), even when the scanned area is as small as 110 × 110 μm$^2$ (i.e., $\frac{\Phi_a}{\Phi_0} \leq \pm 0.15$). Both the vortices are highly extended and intricate, in agreement with the vortex features shown in Fig. 4(b). After acquiring the SSM image shown in Fig. 5(a), we increased the temperature of the system to 40 K and then performed the SSM measurement for the same area [Fig. 5(b)]. The resultant SSM image showed no vortex features, further confirming that these vortices with opposite polarity result neither from any magnetic impurities nor from spurious signals caused by surface roughness. The current distributions calculated for the two vortices shown in Fig. 5(a) are given in Figs. 5(c) and 5(d). For both the vortices, highly diffused and meandering current paths are present around the core region, suggesting elongation of λ. Assuming that $z$ = 6.6 μm, the field distributions of the extended vortices shown in Fig. 5(a) were analyzed by the monopole model, i.e., Eq. (1). The fitted values of λ are 10.4±0.26 and 5.4±0.14 μm for the vortices with negative and positive polarities, respectively. It should be noted, however, that Eq. (1) is applicable only to a thick superconductor in the limit of $\lambda \ll \sqrt{r^2+z^2}$. For the extended vortices observed here, this limit may not be valid, implying that the fitted λ values obtained by the monopole model are only semiquantitative. In order to evaluate the λ values more accurately, we then employed the London model in the thick limit [53,57].

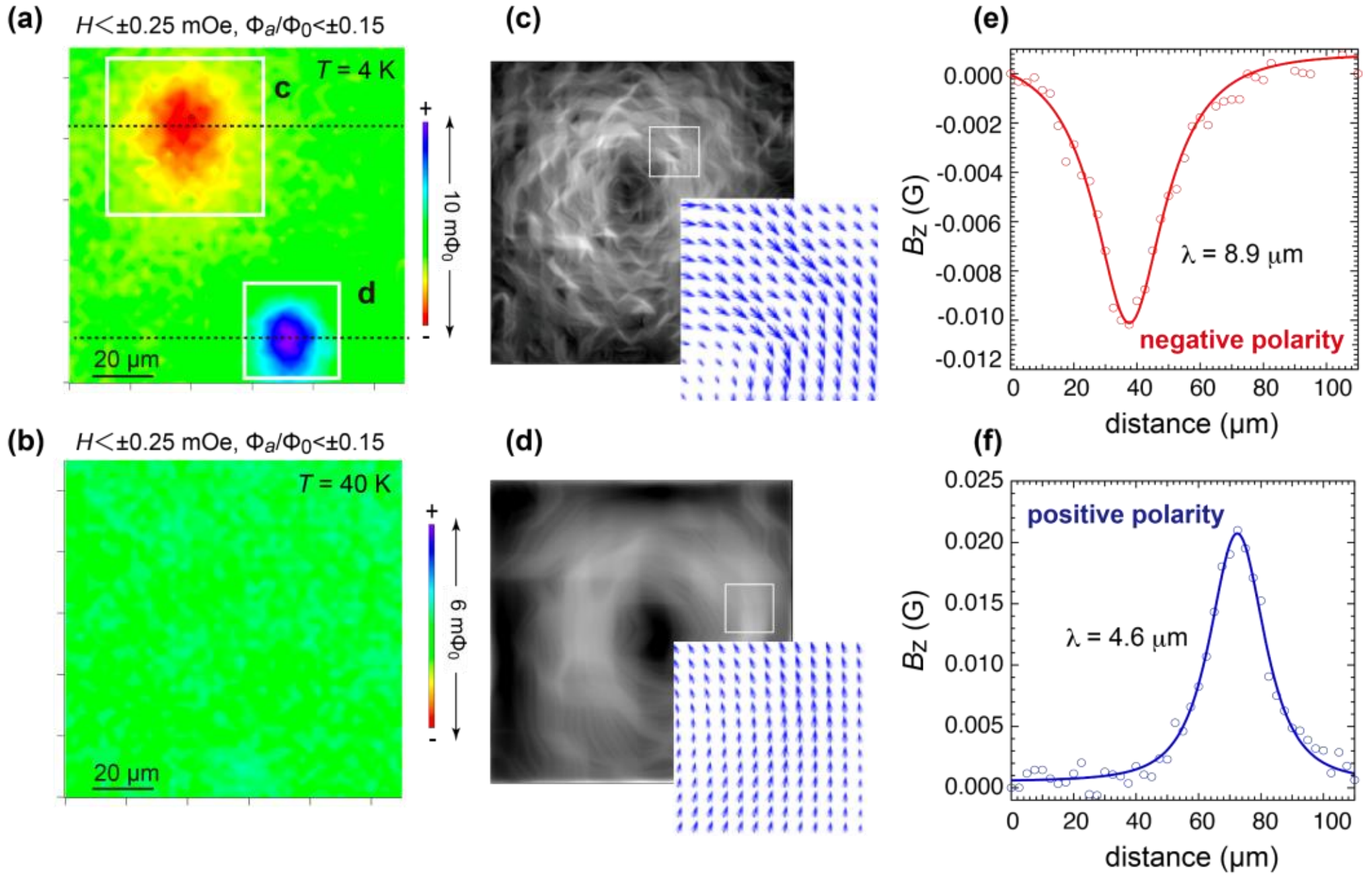

FIG. 5. (a,b) SSM image taken at a temperature of (a) 4 and (b) 40K over an area of 110 × 110 μm$^2$ under nearly zero applied field ($H \leq \pm 0.25$ mOe) conditions with a pixel size of 2.5 × 2.5 μm$^2$. (b, c) Local current distributions of the corresponding white box regions in (a). The insets show the calculated current of the respective white box regions in the form of current vectors. (e,f) The field profiles measured along the dotted lines on the vortices with (e) negative and (f) positive polarities, respectively, shown in (a) The solid lines are fittings based on the London model. The fitted values of penetration depth λ obtained by assuming $z$ = 6.6 μm are shown.

$$B_z(r,z) = \frac{\Phi_0}{2\pi\lambda^2}\int_0^{\infty} \frac{qJ_0(qr)e^{-qz}}{Q(Q+q)}dq, \quad z > 0 \,, (2)$$

where $Q = \sqrt{q^2+\lambda^{-2}}$, and $J_0$ is the zeroth-order Bessel function of the first kind. From the numerical fit of the field profiles to Eq. (2) ($z$ = 6.6 μm), we obtained the λ values of 8.9±0.22 μm [Fig. 5(e)] and 4.6±0.12μm [Fig. 5(f)] for the vortices with negative and positive polarities shown in Fig. 5(a), respectively. During the fitting procedure, the spatial averaging effect due to the finite geometry of the SQUID pickup loop was also incorporated (Appendix B). Thus, the monopole model tends to overestimate the λ value by ~10 %, confirming that the London model is necessary for the quantitative estimate of $\lambda$ for these extended spontaneous vortices (see also Appendix C).

The above argument validates our prediction that the extended vortices arise as a result of the elongation of $\lambda$. It is probable that the superfluid density, $n_s(T) \propto \frac{1}{\lambda^2(T)}$ [58], around the vortices shown in Fig. 5(a) is reduced as compared with the rest of the superconducting area. This suggests that the phase inhomogeneity originally present in the early stage of the phase transition is imprinted as a spatially extended vortex..

### C. Stochasticity in the formation of spontaneous vortices

If vortices are created spontaneously during cooling, the resulting vortices would emerge in a stochastic manner. That is, vortices with different polarities and sizes should be generated at various locations from one quench to another. In order to confirm this, we performed further SSM observations at 4 K over the area of 300 × 200 μm² under near-zero field conditions. Before each SSM observation, the sample was heated above $T_c$ (~50 K) and then cooled down to 4 K to erase the previous history. The results of four consecutive SSM measurements are given in Fig. 6(a). As expected, vortices were found to be generated differently each time in terms of position, size and polarity. Again, the field profiles of all the respective vortices were well reproduced by the London model [Figs. 6(b) and 6(c)]. The estimated λ values for the respective vortices vary in the range from ~1 to ~9 μm. Note, however, that although each vortex does not appear exactly at the same position during each run, vortices tend to emerge in similar regions. For example, a single vortex with negative polarity was constantly observed in the near center of the scanned area. This suggests that spontaneous vortices are preferred to be generated in specific regions where the pinning potential is expected to be strong. At temperatures close to $T_c$, there is practically no pinning, and hence vortices are free to move, disperse and annihilate due to the repulsive (attractive) forces between neighboring vortices with parallel (antiparallel) polarities [59]. Thus, even if vortices are created spontaneously near $T_c$, only the vortices which are effectively immobilized are preserved after further cooling. This argument may also explain why spontaneous vortices (identified as extended vortices) are rarely observed under applied fields, as seen in Fig. 4(a). Under field cooled conditions, the strong pinning sites will be preferentially occupied by field-induced vortices during cooling below $T_c$, and hence spontaneous vortices, if created, might be easily dissipated and/or annihilated before effectively being immobilized.

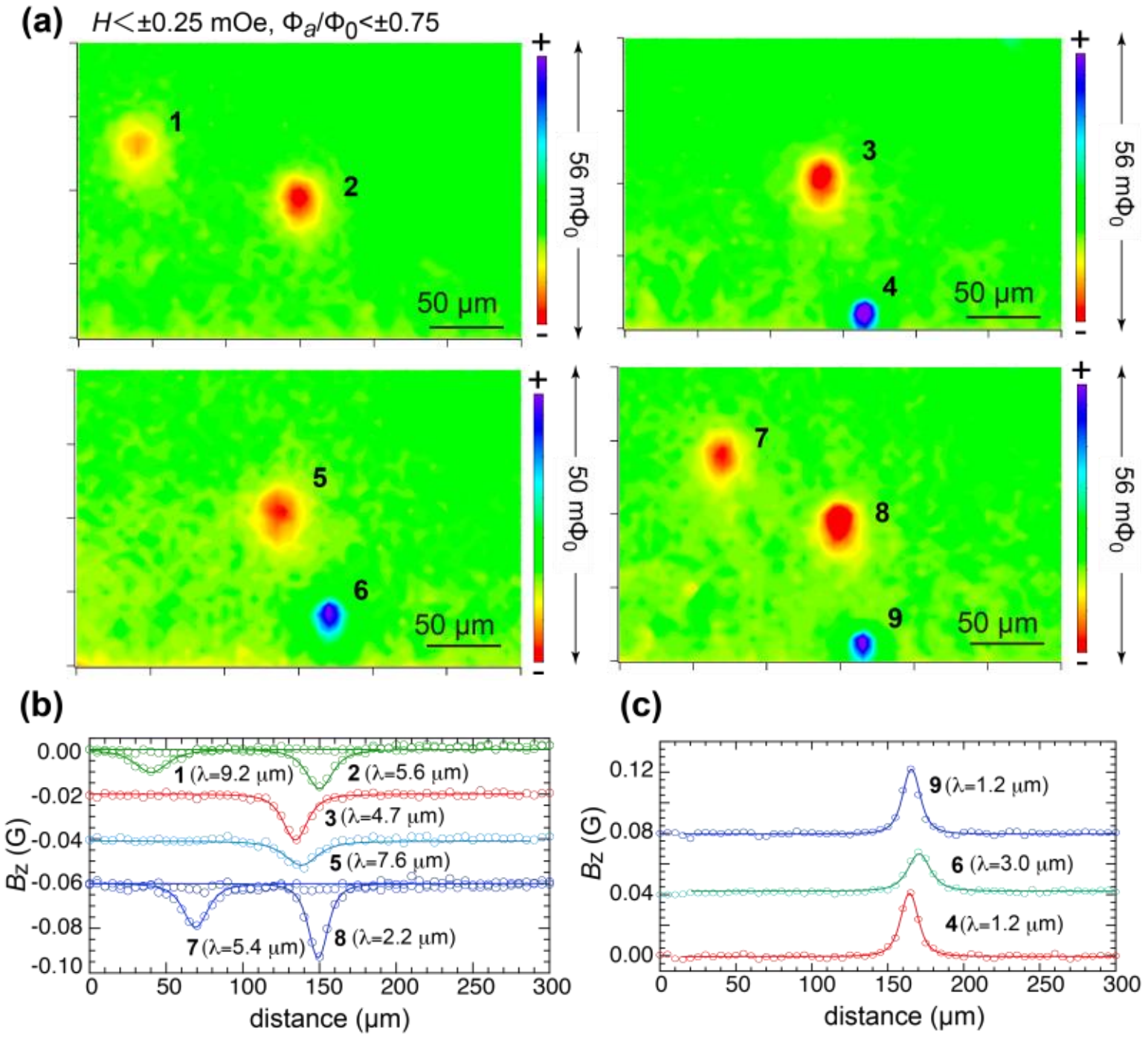

FIG. 6. (a) A series of SSM image taken at a temperature of 4 K over the same area of 300 × 200 μm² under nearly zero applied field ($H$≤±0.25 mOe) conditions with a pixel size of 5 × 5 μm². Each SSM image was taken consecutively after heating the sample up to ~50 K and then cooled down to 4 K to erase the previous history. (b, c) The field profiles measured for the vortices with the same numbering shown in (a). The solid lines are fittings based on the London model ($z$ = 6.6 μm),

## IV. DISCUSSION

Since the λ values of the vortices nucleated under near zero field conditions are in the range of a few to several micrometers, it is interesting to investigate the field distributions of such vortices, for example, a vortex with $\lambda$ = 5 μm, above ($z > 0$), at ($z = 0$), and below ($z < 0$) the surface of the superconductor. According to the London theory, $B_z(r,z)$ and $B_r(r,z)$ for $z \leq 0$ are given by [57],

$$B_z(r,z) = \frac{\Phi_0}{2\pi\lambda^2}\int_0^\infty \left[\frac{q}{Q^2} - \frac{q^2 e^{Qz}}{Q^2(q+Q)}\right] J_0(qr)dq, \quad z \leq 0 \quad (3)$$

$$B_r(r,z) = \frac{\Phi_0}{2\pi\lambda^2}\int_0^\infty \frac{qJ_1(qr)e^{Qz}}{Q(q+Q)}dq, \quad z \leq 0 \quad (4)$$

where $J_n(x)$ is the $n$-th order Bessel function. On the other hand, $B_r(r,z)$ for $z > 0$ is represented by [57]

$$B_r(r,z) = \frac{\Phi_0}{2\pi\lambda^2}\int_0^\infty \frac{qJ_1(qr)e^{-qz}}{Q(q+Q)}dq, \quad z > 0. \quad (5)$$

Figure 7(a) shows streamlines of the magnetic field for a single vortex with $\lambda$= 5 μm calculated numerically by using Eqs. (2)–(5). The effect of the finite vortex core of a radius $\xi$ was considered by inserting a cutoff factor $\exp(-\xi^2 q^2)$ in Eqs. (2) – (5) [57]. We tentatively assumed that the value of $\xi$ is 20 nm to calculate the magnetic field inside and outside of the superconductor. In Fig. 7(a), the magnetic streamlines are plotted as contour lines of magnetic flux with equal intervals. Although this result should be applied, in principle, to a homogeneous superconductor, the present calculations will also provide underlying features of spontaneous vortices near the surface of our sample with $\lambda = 5\ \mu$m. It is clear from Fig. 7(a) that the vortex starts to "open up" even below the surface, in agreement with the vortex spreading feature expected from $\lambda$= 5 μm. We next calculate the perpendicular magnetic field at the surface $B_z(r,0)$ and the corresponding current density $J(r,0)$, which is given by (for the derivation, see Appendix D),

$$J(r,0) = \frac{\Phi_0}{2\pi\mu_0\lambda^2}\int_0^\infty \frac{qJ_1(qr)}{Q}dq, \quad (6)$$

where $\mu_0$ is the permeability of free space. A one-dimensional representation of our results for a single vortex with $\lambda$= 5 μm is shown in Fig. 7(b). One sees from Fig. 7(b) that substantial perpendicular field and the resultant superconducting current exist on a length scale of several micrometers around the core. With this in mind, we illustrate in Fig. 7(c) the possible vortex supercurrent distribution over the surface by superimposing on a typical scanning electron microscope (SEM) image of the present sample. In the SEM image, the gray and black regions correspond to the MgO- and $MgB_2$-rich regions, respectively, demonstrating the fractal and interwoven-like $MgO/MgB_2$ structures in the sub-micrometer length scale. As illustrated in Fig. 7(c), the vortex supercurrents possibly circulate over a number of $MgO/MgB_2/MgO$ junctions. We have confirmed from cathode-luminescence imaging that in the MgO-rich regions, there exist channels of oxygen vacancies that surround the $MgB_2$-righ regions [38] (see also Fig. S2 in Supplemental Material [39]). These channels of oxygen vacancies provide conductive paths for both electrons and hole currents [60,61], resulting not only in the metal-like conductivity in the MgO-rich region, but also in consecutive Andreev reflections over several $MgO/MgB_2$ interfaces [see also the inset of Fig. 7(c)]. The resulting vortex current will lead to the meandering current paths, as shown in the red line in Fig. 7(c). Thus, the current distribution expected from the local spatial morphology of the MgO and $MgB_2$ grains is fully consistent with the SSM image and the related current distribution of the vortices formed under near-zero field conditions shown in Fig. 5. We therefore infer that the original scenario of spontaneous vortex formation in a proximity-coupled nanocomposite (see Fig. 1) is consistent with the present MgO/$MgB_2$ nanocomposite.

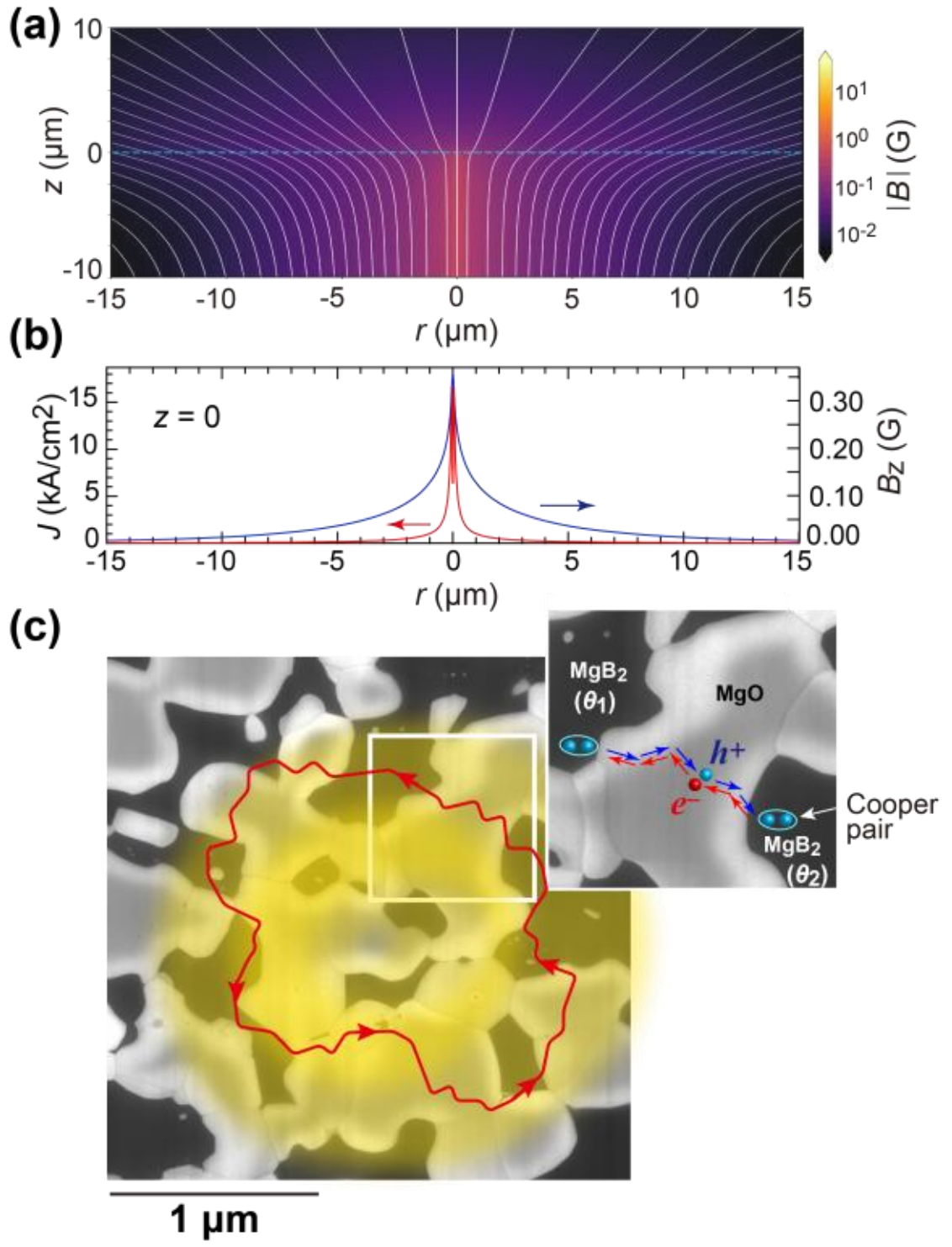

FIG. 7. (a) Magnetic field lines of a straight vortex with $\lambda$= 5 μm in a superconducting half space calculated from the London model. (b) Perpendicular magnetic field $B_z(r, 0)$ and the corresponding current density $J(r, 0)$ at the surface of the superconductor. (c) Expected vortex current distribution (yellow region) superimposed on a typical scanning electron microscope (SEM) image of the present nanocomposite. A red line indicates a possible current loop. The inset illustrates a schematic representation of the Andreev reflection process occurring over the $MgB_2$/MgO/$MgB_2$ junction enlarged from the white box region.

Finally, we would like to comment on why vortices can be created spontaneously in our nano-composite even under a relatively slow cooling rate (~0.2 K/s). In the temperature region just below $T_c$, each $MgB_2$-rich domain in the nanocomposite will immediately reach internal equilibrium under such slow cooling conditions. The respective $MgB_2$-rich domains are uncoupled as long as the coupling energy of the junction, $E_J \equiv \frac{I_c \Phi_0}{\pi}$ [20], is smaller than the thermal energy $k_\mathrm{B}T$, where $I_c$ is the maximum junction current and $k_\mathrm{B}$ is Boltzmann's constant. As the temperature is cooled further, $I_c$ and hence $E_J$ will increase accordingly. Eventually, $E_J$ will become large enough to couple the neighboring $MgB_2$ domain. This can lead to the formation of coherent closed loop consisting of several numbers ($N$) of MgO/$MgB_2$/MgO junctions. When the sum of the phase differences in a loop equals to $2\pi$ (or $-2\pi$) and the total coupling energy of the loop $NE_J$ becomes larger than the vortex energy needed to support $\Phi_0$, a vortex with one flux quantum can be nucleated. From this argument follows that larger loops are more energetically preferable because of the larger values of $NE_J$, possibly accounting for the formation and low-temperature stability of spatially extended vortices with $\lambda$ greater than several micrometers. According to the above scheme, spontaneous vortices are created as a result of the competition between the Josephson coupling energy and the thermal phase fluctuation. In the original KZ model, the cooling rate must be fast enough to freeze the value of the phase difference in each superconducting domain whose typical size is determined by the freeze-out coherence length $\xi$ [1,18]. In our case, on the other hand, $MgB_2$-rich domains with sizes of sub-micrometers to micrometers already reach internal equilibrium independently and choose their own phase at the time when the coherent loop is established. Thus, the fast cooling is not required to retain the phase difference. This situation is similar to the case of a multi-junction-Josephson loop consisting of hundreds of grain-boundary Josephson junctions connected in series [22]. In our case, however, the size and shape of spontaneous vortices can be monitored directly by an appropriate magnetic imaging technique, such as SSM, revealing unique geometrical characteristics of the spontaneous vortices in a proximity-induced superconductor. Further analysis on spontaneous vortices using, for example, state-of-the-art SQUID-on-tip probes [62,63] will be useful for a better and comprehensive understanding of the spontaneous vortices nucleated in this highly disordered proximity-coupled system.

## IV. CONCLUSIONS

We have imaged and characterized spontaneous vortices generated in a proximity-induced MgO/$MgB_2$ nanocomposite using scanning SQUID microscopy. In contrast to field-induced Abrikosov vortices, the spontaneous vortices nucleated under near-zero-fields exhibit highly extended and intricate topologies. Quantitative fitting based on the London model revealed anomalously long penetration depths varying from ~1 to ~9 μm. Our measurements have further demonstrated that these vortices emerge stochastically in terms of position, size, and polarity. Since the micrometer- to sub-micrometer-sized $MgB_2$ domains are structurally predefined and isolated from each other in the highly disordered fractal matrix, they choose their own phase independently at temperatures near $T_c$. The subsequent formation of global phase coherence links these independent phases together through a network of strongly coupled proximity junctions. When the phase accumulation around a newly established loop meets topological boundary conditions, a single flux quantum becomes frozen to yield a geometrically extended vortex, which continues to survive even after the system is cooled further. The meandering supercurrent profiles serve as a frozen structural footprint of the sample's local phase fluctuations in the early stage of the phase transition. This mechanism implies that ultra-fast thermal quenching is not necessarily required for topological defect formation in highly disordered proximity-coupled systems. Thus, the proposed defect formation mechanism can be regarded as "geometrically driven," and the resulting vortices are "topologically protected" due to the potentially large total coupling energy of the extended loop. Taken together, these findings illustrate how spontaneous defect formation in an inhomogeneous superconductor can differ fundamentally from that in a homogeneous one. They provide experimentally accessible insights into the structure, formation, and stabilization of topological defects in highly disordered superconducting systems, whose physical understanding is still a subject of ongoing research [64–66]. Our results therefore highlight the broader relevance of proximity-coupled and structurally complex superconductors as model platforms for exploring phase ordering and defect nucleation in systems dominated by fractal disorder and the related hierarchical inhomogeneity.

## ACKNOWLEDGMENTS

A part of this work was conducted in Institute for Molecular Science, supported by "Advanced Research Infrastructure for Materials and Nanotechnology in Japan (ARIM)" of the Ministry of Education, Culture, Sports, Science and Technology (Proposal Numbers. JPMXP1223NM0163, JPMXP1223MS1024 and JPMXP1224MS1017). This work was also supported by NIMS Joint Research Hub Program. We also acknowledge Research Facility Center for Science and Technology, Kobe University, for providing access to the MPMS facility.

## DATA AVAILABILITY

The SSM data and analysis routines (Jupyter/Python) needed to reproduce the figures can be found in Ref. [67].

## APPENDIX A: MONOPOLE APPROXIMATION

The magnetic field profile of the vortex's z-component $B_z(x,y,z)$ of the stray field sufficiently above a superconductor can be approximated by the stray field of a magnetic monopole. For a thick sample with thickness $d$ ($d \gg \lambda$) and in the limit of $\sqrt{x^2+y^2+z^2} = \sqrt{r^2+z^2} \gg \lambda$, $B_z(r,z)$ can be expressed as [51,53–56]

$$B_z(r,z) = \frac{\Phi_0}{2\pi} \frac{z+\lambda}{[r^2+(z+\lambda)^2]^{\frac{3}{2}}}, \quad z > 0, \quad \text{(A1)}$$

where $r = (x,y)$ is the distance from the vortex center, $\lambda$ is the magnetic penetration depth, $z$ is the distance from the sample surface to the SQUID sensor. In scanning magnetic imaging measurements, the accuracy of the monopole model is improved by averaging over an area $S$ representative of the sensing probe [55,56],. The area of the loop with radius $r_0$ is obtained by the following integration:

$$S = \int_{-r_0}^{r_0} 2\sqrt{r_0^2 - r'^2}\, dr'. \quad \text{(A2)}$$

Hence, the measured $B_z(r,z)$ is well represented by the convoluted value $\langle B_z(r,z)\rangle$ over the sensing area of the SQUID pick-up loop through

$$\langle B_z(r,z)\rangle = \frac{\Phi_0}{2\pi S}\int_{r-r_0}^{r+r_0} 2\sqrt{r_0^2-(r'-r)^2}\, \frac{z+\lambda}{[r^2+(z+\lambda)^2]^{\frac{3}{2}}}\, dr'. \quad \text{(A3)}$$

In our measurements, a SQUID pick-up loop with an active area of $S = \pi r_0^2$ ($r_0$=5 μm) was used. It should also be noted that as for the field-induced Abrikosov vortices in the present MgO/$MgB_2$ nanocomposite, $\lambda$, which is less than 100 nm at sufficiently low temperatures [36,37], is much smaller than $z$, which is in the order of several micrometers. Thus, we can safely approximate $z + \lambda \approx z$. Consequently, we can evaluate the value of $z$ by fitting the field profile of the vortices in the SSM images taken under field cooling at 4 K to Eq. (A3). For this purpose, we selected six isolated vortices with different polarities (see Fig. S5 in Supplemental Material [39]), and the $z$ value was determined to be 6.6±0.3 μm.

## APPENDIX B: LONDON MODEL

For the vortices with λ greater than a few micrometers, the limit of $\sqrt{r^2+z^2} \gg \lambda$ is not valid anymore, and hence the full London model [53,57] should be used. In dealing with the convolution effect of the SQUID pick-up loop, we used a different method than that employed in Appendix A to avoid double integrals. From the exact solution to the London model with continuous boundary conditions at the superconductor-vacuum interface, the vertical magnetic field component $B_z(r,z)$ in the vacuum region ($z$>0) [see Eq. (2)] is transformed into Fourier-Hankel space ($q$),

$$\widetilde{B_z}(q,z) = \frac{\Phi_0}{2\pi\lambda^2}\frac{e^{-qz}}{Q(Q+q)}, \quad \text{(B1)}$$

where $Q = \sqrt{q^2 + \lambda^{-2}}$ . We define the loop's geometric sensitivity profile using a top-hat aperture function $M(\rho)$:

$$M(\rho) = \begin{cases} 1 & \rho \le r_0 \\ 0 & \rho > r_0 \end{cases} \quad \text{(B2)}$$

To solve the convolution in Hankel space, we take the 2D transform of this circular aperture:

$$\widetilde{M}(q) = \int_0^{r_0} \rho J_0(q\rho)d\rho = \frac{r_0 J_1(qr_0)}{q}. \quad \text{(B3)}$$

By applying the Hankel convolution theorem, the total magnetic flux $\Phi(r,z)$ passing through the pickup loop with radius $r_0$ is calculated by integrating the product of $\widetilde{B_z}(q,z)$ , $\widetilde{M}(q)$ , and a phase-shifting Bessel function $J_0(qr)$:

$$\begin{aligned}\Phi(r,z) &= \int_0^\infty \int_0^{2\pi} q\,\widetilde{B_z}(q,z)\widetilde{M}(q)J_0(qr)d\theta dq \\ &= 2\pi\int_0^\infty q\,\widetilde{B_z}(q,z)\widetilde{M}(q)J_0(qr)dq \\ &= \frac{\Phi_0 r_0}{\lambda^2}\int_0^\infty \frac{J_1(qr_0)J_0(qr)e^{-qz}}{Q(Q+q)}dq. \quad \text{(B4)}\end{aligned}$$

Considering that convoluted value $\langle B_z(r,z)\rangle$ is represented by

$$\langle B_z(r,z)\rangle = \frac{\Phi(r,z)}{\pi {r_0}^2}, \quad \text{(B5)}$$

we reach the following formula for calculating $\langle B_z(r,z)\rangle$:

$$\langle B_z(r,z)\rangle = \frac{\Phi_0}{\pi r_0 \lambda^2}\int_0^\infty \frac{J_1(qr_0)J_0(qr)e^{-qz}}{Q(Q+q)}dq. \quad \text{(B6)}$$

As mentioned in Appendix A, the z value was determined to be 6.6$\pm$0.3 μm. Once the z value is known, we can obtain the value of $\lambda$ for spatially extended vortices (such as those shown in Figs. 5 and 6) by fitting their field profile to the equation (B6). The values of $\lambda$ obtained under the condition $z$ = 6.6 μm are given in Figs. 5(e), 5(f), 6(b) and 6(c).

## APPENDIX C: COMPARISON BETWEEN LONDON MODEL AND MONOPOLE MODEL

It is instructive to investigate how the distribution of the magnetic field lines calculated by the London model or the monopole model varies depending on the value of λ. Figure 8(a) shows the magnetic field lines of a straight vortex with λ = 1, 5 or 10 μm in a superconducting half space calculated numerically by the London model, i.e. by using Eqs. (2)–(5). The effect of the widening of the vortex field in a superconducting half space becomes more evident with increasing λ. In the monopole model, on the other hand, the effect of field spreading below the surface is neglected,

$$B_z(r,z) = \frac{\Phi_0}{2\pi\lambda^2}\int_0^\infty \frac{q}{Q^2}J_0(qr)dq\,.\ z \le 0. \quad \text{(C1)}$$

Also, in the monopole model or in the limit of $\sqrt{r^2+z^2} \gg \lambda$, the perpendicular component $B_z(r,z)$ of the stray field above a superconductor is represented by Eq. (A1), and the corresponding horizontal component $B_r(r,z)$ is given by

$$B_r(r,z) = \frac{\Phi_0}{2\pi}\frac{r}{[r^2+(z+\lambda)^2]^{\frac{3}{2}}}, z > 0. \quad \text{(C2)}$$

The resulting magnetic field lines calculated numerically from Eqs. (A1), (C1) and (C2) are shown in Fig. 8(b). Figure 8(c) compares the $B_z(r,z)$ profiles of the vortex with λ = 1, 5 or 10 μm at $z$ = 6.6 μm obtained by these two models. As for the vortex with λ = 1 μm, the field profiles, with full width at half maximum (FWHM) of ~10 μm, are almost identical between the two models. For the vortices with λ = 5 and 10 μm, the monopole model yields larger values of $B_z(r,z)$ especially in the region of smaller $r$, as compared with the London model. These results allow us to confirm that the monopole model is not valid for the vortices with λ greater than a few micrometers. Stated in another way, when the field profile of a vortex is characterized by the FWHM of more than ~20 μm in our experimental set-up, their λ values will be large enough to violate the monopole approximation.

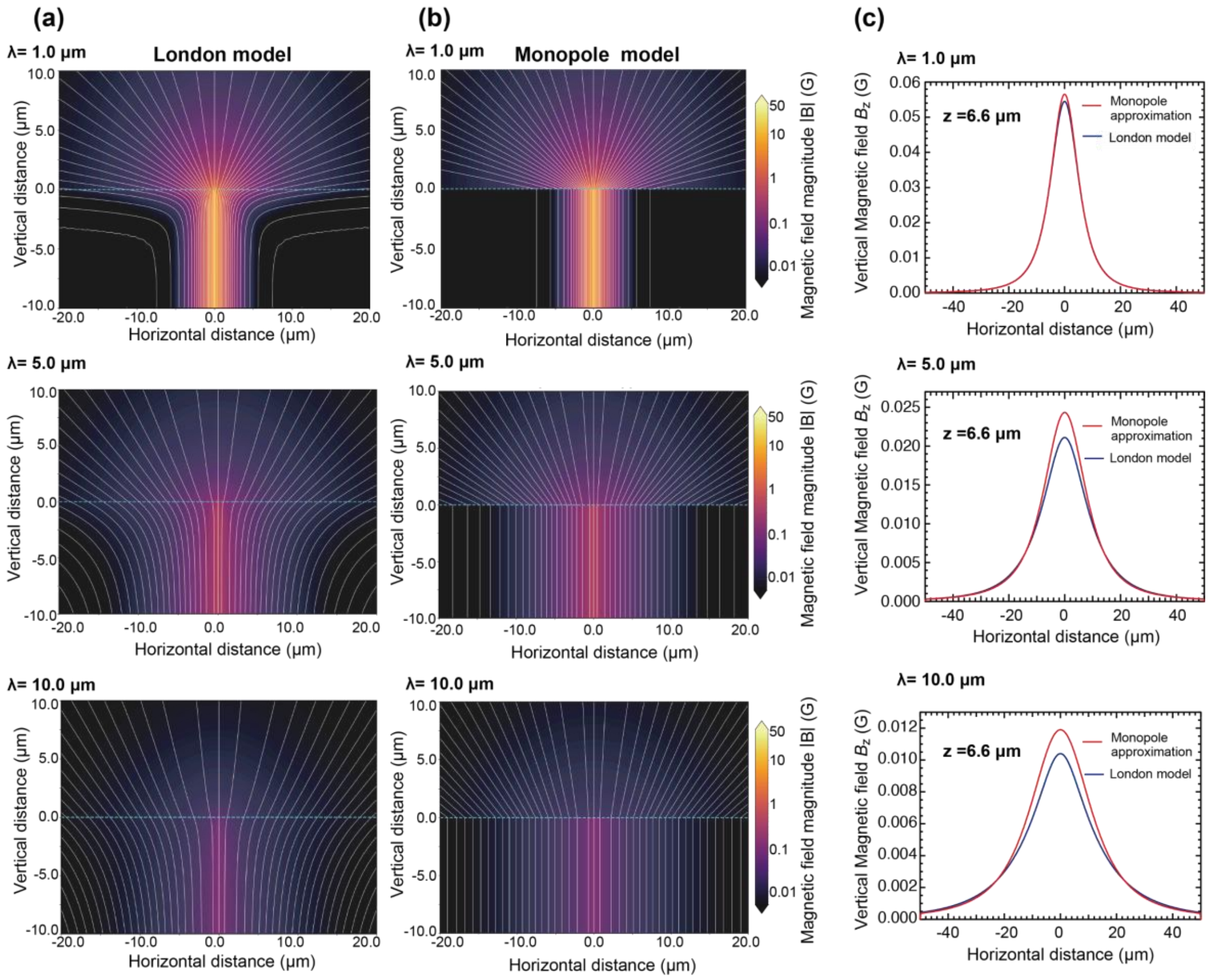

FIG. 8. (a,b) Magnetic field lines of a straight vortex with different values of λ [λ= 1 μm (upper), 5 μm (middle) and 10 μm (lower)] in a superconducting half space calculated from (a) the London model and (b) the monopole model. (c) The corresponding profiles of the perpendicular magnetic field $B_z$ at $z$ = 6.6 μm.

## APPENDIX D: INTEGRATION KERNEL OF SURFACE CURRENT DENSITY

According to Ref. [57], the current density $J(r,z)$ derived from a straight vortex in a superconducting half space ($z \leq 0$) is given by,

$$J(r,z) = \mu_0\lambda^{-2}\left[\frac{\Phi_0}{2\pi r} - A(r,z)\right], \quad \text{(D1)}$$

$$A(r,z) = \frac{\Phi_0}{2\pi\lambda^2}\int_0^\infty \frac{J_1(qr)}{Q^2} f(q,z)dq, \text{(D2)}$$

$$f(q,z) = 1 - \frac{q}{q+Q}e^{Qz}. \text{(D3)}$$

From $\frac{1}{r} = \int_0^\infty J_1(qr)dq$, Eqs. (D1)–(D3) are reduced to the following equation:

$$J(r,z) = \frac{\Phi_0}{2\pi\mu_0\lambda^2}\int_0^\infty \left[\frac{q^2}{Q^2} + \frac{q}{\lambda^2 Q^2(Q+q)}e^{Qz}\right]J_1(qr)dq. \text{(D4)}$$

Setting z = 0 in Eq. (D4), the surface current density $J(r,0)$ is given by,

$$J(r,0) = \frac{\Phi_0}{2\pi\mu_0\lambda^2}\int_0^\infty \left[\frac{q^2}{Q^2} + \frac{q}{\lambda^2 Q^2(Q+q)}\right]J_1(qr)dq.$$

Since $\lambda^2 = (Q^2 - q^2)^{-1}$ from $Q = \sqrt{q^2 + \lambda^{-2}}$, we finally obtained the following simple form to numerically calculate $J(r,0)$,

$$J(r,0) = \frac{\Phi_0}{2\pi\mu_0\lambda^2}\int_0^\infty \frac{qJ_1(qr)}{Q}dq. \text{(D5)}$$

# Supplemental Material for

## Imaging and characterization of spontaneous vortices in a proximity-induced superconductor

Iku Nakaaki,[1] Kotaro Taki,[1] Mio Nomura,[1] Haruna Ishimaru,[1] Minagi Yono,[1] Jun Chen,[2] Hiroyo Segawa,[2] Akiko Nakamura,[3] Taku Moronaga,[3] Minoru Tachiki,[4] Shuuichi Ooi,[4] Shunichi Arisawa,[5] Tsutomu Nojima,[6] and Takashi Uchino[1]

[1]*Department of Chemistry, Graduate School of Science, Kobe University, Nada, Kobe 657-8501, Japan*

[2]*Research Center for Electronic and Optical Materials, National Institute for Materials Science, Tsukuba, Ibaraki 305-0044, Japan*

[3]*Research Network and Facility Services Division, National Institute for Materials Science, Tsukuba, Ibaraki 305-0047, Japan*

[4]*International Center for Materials Nanoarchitectonics (MANA), National Institute for Materials Science, Tsukuba, Ibaraki 305-0047, Japan*

[5]*Research Center for Functional Materials, National Institute for Materials Science, Tsukuba, Ibaraki 305-0047, Japan*

[6]*Institute for Materials Research, Tohoku University, Sendai, Miyagi 980-8577, Japan*

## Methods

### Sample preparation.

The bulk Mg/MgO/$MgB_2$ nanocomposites were prepared by the solid phase reaction of Mg and $B_2O_3$ powders under Ar atmosphere at 700 ˚C followed by a subsequent spark plasma sintering (SPS) procedure according to the procedure reported in our previous papers[36-38]. Pure Mg (99.9%) and $B_2O_3$ (99.9%) powders with a molar ratio of Mg/$B_2O_3$=5:1 were used as starting material. The Mg/$B_2O_3$ mixture with a total weight of 2 g was put in a cylindrical alumina crucible. This crucible was located inside a larger rectangular alumina crucible, which was closed with a thick aluminum lid. This set of crucibles was placed in an electric furnace, which was evacuated down to ~30 Pa and was subsequently purged with argon. The temperature of the furnace was raised to 700 ˚C at a rate of ~10 ˚C/min and kept constant at 700 ˚C for 6 h under flowing Ar environment. After the heating process, the furnace was naturally cooled to room temperature, yielding black powders in the inner crucible. We confirmed from powder X-ray diffraction (XRD) measurements that the resulting powder consists of MgO, Mg, and $MgB_2$. The collected black powders weighing about 2.0 g were loaded into a 15 mm diameter graphite die and were processed using a SPS system (SPS-322HG, Fuji Electronic Ind. Co., LTD.). In SPS, sintering is realized by subjecting the green compact to arc discharge generated by a pulsed electric current. An electric discharge process takes place on a microscopic level and accelerates the sintering processes accompanied by material diffusion. The pulsed electric current (2000 A) was passed through the sample under dynamic vacuum (~50 Pa) while a 114 MPa uniaxial pressure was applied. The heating rate was ~450 ˚C/min up to 1200 ˚C. During the SPS procedure, the temperature, applied pressure, displacement (shrinkage), and environmental pressure were recorded. After the SPS treatment, a cylindrical bulk sample with a diameter of 15 mm and a length of ~4 mm was obtained. The bulk density of 3.15 g/cm$^3$, which corresponds to 97 % densification, was obtained by dividing the weight in air by the geometric volume of the specimen. The thus obtained SPS-treated samples were cut into suitable shapes and then mirror-polished using diamond paste for later characterization.

As a reference, we also prepared a pure $MgB_2$ bulk sample by a similar SPS method using commercially available $MgB_2$ powder (Sigma Aldrich, ≥99% trace metals basis, -100 mesh).

### X-ray diffraction measurements.

X-ray diffraction (XRD) patterns of the SPS-treated sample were obtained with a diffractometer (SmartLab, Rigaku) equipped with a sealed tube X-ray generator (a copper target; operated at 40 kV and 30 mA). Rietveld refinement of XRD patterns was performed to accurately determine the phases and their quantitative compositions using Rigaku's PDXL software with the Whole Pattern Powder Fitting (WPPF) method, connected to the PDF2 database of International Centre for Diffraction Data (ICDD). The refinements were carried out by checking the fitting quality of the powder patterns by means of two parameters, i.e., the goodness-of-fit indicator GoF and the weighted R-factor of Rietveld refinement $R_{wp}$.

**Focused ion beam scanning electron microscopy (FIB-SEM) observation.**

FIB and energy dispersive X-ray (EDX) spectroscopy were conducted on the sample cut from the SPS-treated samples using an FIB-SEM system (SMF-1000, HITACHI High-Tech) with an EDX spectrometer. The specimen was milled using a Ga ion beam at a 30 kV acceleration voltage and a 3 nA beam current with a milling interval of 25.0 nm. The milled surfaces of a $5 \times 5$ μm$^2$ region with a pixel size of $2.5 \times 2.5$ nm$^2$ were imaged by SEM at an acceleration voltage of 5 kV using in-lense SE detector. From this procedure, we obtained a set of 1000 FIB-SEM serial sectioning images. After aligning the data set, image stacks were segmented to reconstruct a 3D image with a voxel size of $2.5 \times 2.5 \times 25.0$ nm$^3$.

**Fractal analysis.**

To evaluate the fractal dimension for the distribution of $MgB_2$ nanograins in the composite, box counting was applied on SEM images. Firstly, the original images were converted to an 8-bit format and then finally transformed to a binary format. Image processing and fractal analysis were carried out in the ImageJ 1.52a software (NIH). Box counting technique is consisted of covering the slice image with voxels (boxes in 2D). The fractal dimension or box-counting dimension $D$ is obtained from the following relationship [40]

$$D = \frac{\log(N)}{-\log(r)}, \qquad (S1)$$

where $r$ is variable box dimension in pixels and $N$ is the minimum number of boxes needed to encompass the whole object containing the boron K□□signals. If any linear region is observed in the curve $\log(N)$ versus $\log(r)$, the fractal dimension is equal to the slope with a negative sign of the curve.

**Electrical resistivity and current-voltage (*I-V*) measurements.**

The electrical resistivity of the square cuboid sample with a size of 1×2×10 mm$^3$ was measured using the standard dc four-probe technique in the 2□300 K range. For measuring the *I-V* characteristics, a special combined mode (delta-mode) operation of the current source (6220, Keithley Instruments, Cleveland, USA) and nanovoltmeter (2182A. Keithley Instruments, Cleveland, USA) was used. Constant current pulses with reversing polarity were sent through the sample and the voltages were measured at two intermediate points along the length. Using the short reversing current pulses, resistive heating in thin samples was avoided, and the spurious Seebeck voltages that may cause erroneous electrical resistivity results were also eliminated.

**DC and AC magnetization measurements.**

The dc and ac magnetization measurements in external fields $H$ were carried out for a square cuboid shape sample with dimensions of 1×1×5 mm$^3$ using a commercial SQUID magnetometer (MPMS-XL and MPMS-3, Quantum Design, San Diego, USA). The field was applied along the long side of the cuboid sample. In this configuration, the resulting effective demagnetization factor is estimated to be less than 0.05 [41], and the magnetizations and susceptibilities were not corrected for this small demagnetization factor. To estimate the volume magnetization $M$, we used the bulk density of the sample (3.15 g/cm$^3$). The magnetic critical current density $J_c(H)$ in A/cm$^2$ was determined using the Bean critical state model for a rectangular sample:

$$J_c(H) = \frac{20\Delta M}{\left(a - \frac{a^2}{3b}\right)}, \qquad (S2)$$

where $a$ and $b$ ($a \leq b$ in cm) are cross-sectional sample-dimensions perpendicular to the magnetic field, and □$M$ is the difference between the upper and lower branches of the hysteresis loops in emu/cm$^3$.

**Local current distribution obtained by an inversion of Biot-Savart law.**

The calculation of current components from the measured perpendicular component of the magnetic field $B_z$ is possible by using an inversion of Biot-Savart law. Several attempts to solve this problem for 1-dimensional (1D), 2D and 3D samples were reported [42,43]. As for the problem in 2D, a general solution was developed by Jooss *et al*. [44] and has been recently refined by Zuber *et al*. [45], as given by the following expression:

$$\tilde{J}_x\left(k_x, k_y\right) = \frac{ik_y}{\mu_0} e^{kh} \text{cosech}\left(\frac{kd}{2}\right) \tilde{B}_z\left(k_x, k_{y,}, h, d\right), \quad \text{(S3)}$$

$$\tilde{J}_y\left(k_x, k_y\right) = \frac{-ik_x}{\mu_0} e^{kh} \text{cosech}\left(\frac{kd}{2}\right) \tilde{B}_z\left(k_x, k_{y,}, h, d\right), \quad \text{(S4)}$$

where $k_x$ and $k_y$ are Fourier space wave-vectors, $k = \sqrt{k_x^2 + k_y^2}$, $\tilde{B}_z$ and $\tilde{J}_{x,y}$ are magnetic field and sheet current in Fourier space, respectively, $h$ is the distance between the top of the sample and the bottom of the indicator film, $d$ is the thickness of the sample, and $\mu_0$ is the permeability of free space. Accordingly, two current components $J_x$ and $J_y$ can be obtained by inverse Fourier transformation of equations (S3) and (S4). Here, it should be noted that equations (S3) and (S4) hold exactly for very thin samples, i.e., $h \gg d$, in which the perpendicular component of the current density $J_z$ does not contribute to $B_z$. However, even for thick samples, the measured $B_z$ is mainly sensitive to the surface currents within a surface layer of thickness of ~10 μm [46]. Thus, the 2D Fourier transformation based on the inverted Biot-Savart law can also be used to calculate, although semi-quantitatively, the current distribution in a near-surface layer of thick samples. In this work, we used the values of 3 and 6 μm for $h$ and $d$, respectively, in equations (S3) and (S4) to obtain $J_x$ and $J_y$. We confirmed that the pattern of the current distribution is not sensitive to the particular choice of $h$ and $d$ for as long as $h < 8$ μm and $d < 10$ μm. We therefore consider that the thus obtained current distribution pattern will represent the underlying nature of the vortex current flow in the near-surface region of the sample.

**Cathodoluminescence (CL) measurements.**

CL spectra and images were recorded using a cathodoluminescence spectrometer system (MP32 CL, Horiba) attached to a field emission scanning electron microscope (SU6600, Hitachi). The acceleration voltage was set to 5 or 10 kV, and the probe current was 3.6 or 7.2 nA according to the measurement requirements. The CL grating is 100 grooves/mm in order to cover a wide wavelength range (300–900 nm), and all measurements were conducted at room temperature.

**Scanning SQUID microscopy (SSM) imaging.**

SSM measurements were conducted with a scanning SQUID microscope (SQM-2000, SII Nanotechnology) with a dc SQUID magnetometer consisting of Nb/Al-$AlO_x$/Nb Josephson junctions and an inductively coupled pick-up loop of a Nb film [47]. We used the pick-up loop of 10 μm diameter with 2 μm linewidth. The dc SQUID and the pick-up loop were integrated on a square Si chip of 3 mm×3 mm in size, which was mounted on a phosphor bronze cantilever. The sensor chip was tilted with respect to the sample stage by a shallow angle of ~10 degrees. A magnetic flux resolution is better than 5 $\mu\Phi_0/\text{Hz}^{1/2}$, corresponding to a field noise of $\sim 10^{-10}$ $\text{T/Hz}^{1/2}$ referred to the pickup loop. A corner of the chip was slowly approached to the sample until a specific distance (~5-7 μm) between the loop and the sample surface was achieved. The total spatial resolution of the present measurement system is ~2.5 μm [47]. Although the whole assembly including the sensor chip and the sample stage was surround by a μ-metal shield, a residual magnetic field (ambient field) of ~15 mOe was present. Further cancellation was achieved by applying a small current through a coil wound in the sample stage, reaching a field of about $\pm$0.25 mOe. Near-zero field measurements were performed under these conditions ($H<\pm$0.25 mOe). For field cooled measurements, a small magnetic field in the range from -30 to 30 mOe was applied perpendicular to the sample surface. We scanned the sample stage in *xy* directions by dc stepper motors. During the measurements, the temperature of the SQUID sensor was kept to ~4 K, whereas the sample temperature was varied from 4 to 50 K.

## Supplemental Figures

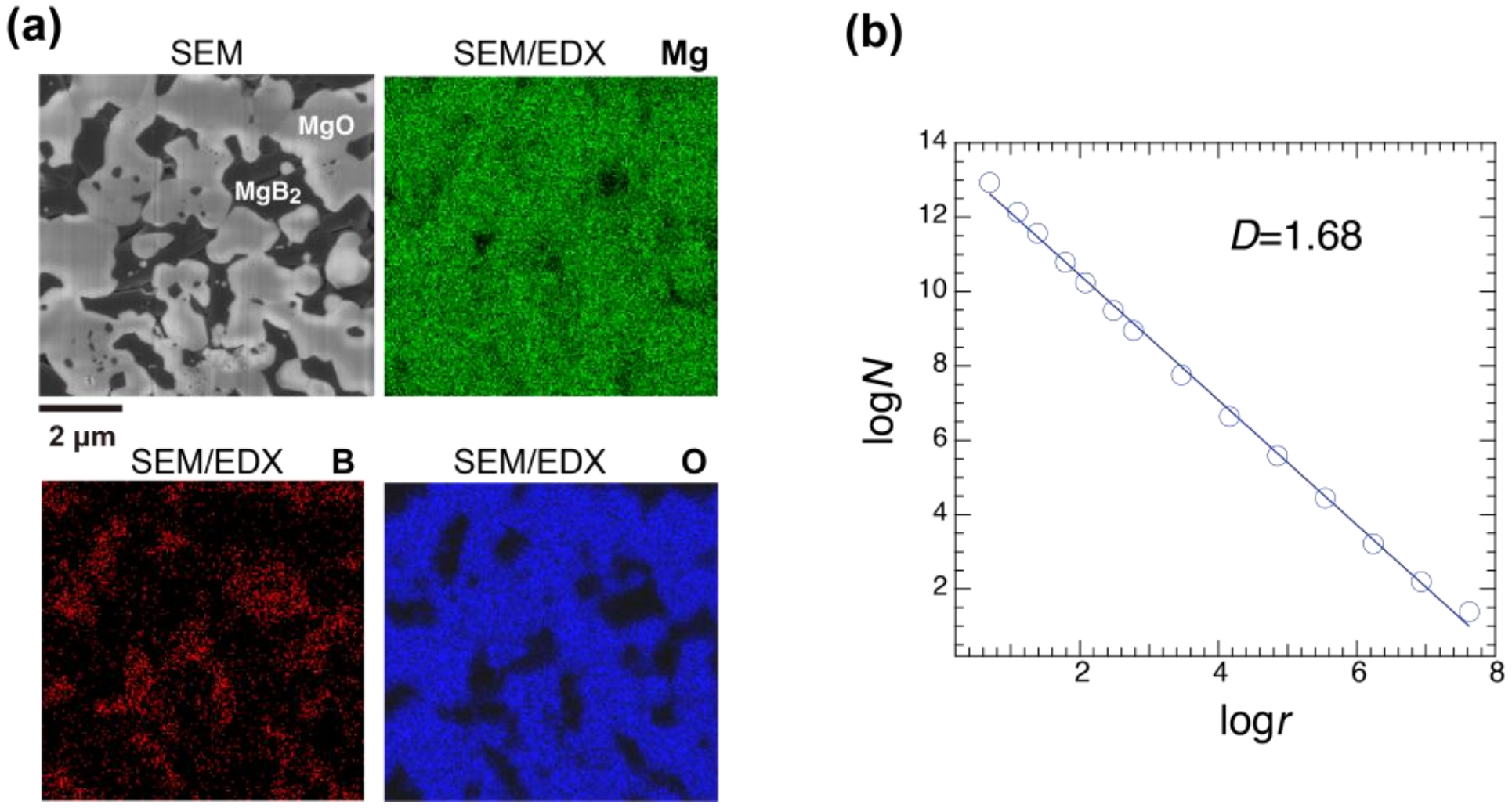

FIG. S1. Fractal dimension of the $MgO/MgB_2$ nanocomposite. (a) Scanning electron microscope (SEM) and the corresponding energy dispersive x-ray spectroscopy (EDX) images of the surface of the $MgO/MgB_2$ nanocomposite. Green = Mg, Red = B, and Blue = O. The SEM-EDX mapping images revealed that the oxygen (blue) and boron (red) distributions almost match with the gray and black regions in the corresponding SEM image. From this follows that the gray and black regions correspond to the MgO and $MgB_2$ rich regions, respectively. (b) The box counting analysis (for details see Methods) for the binarized SEM image given in (a). The box counting fractal dimension *D* can estimated from the slope of the plot.

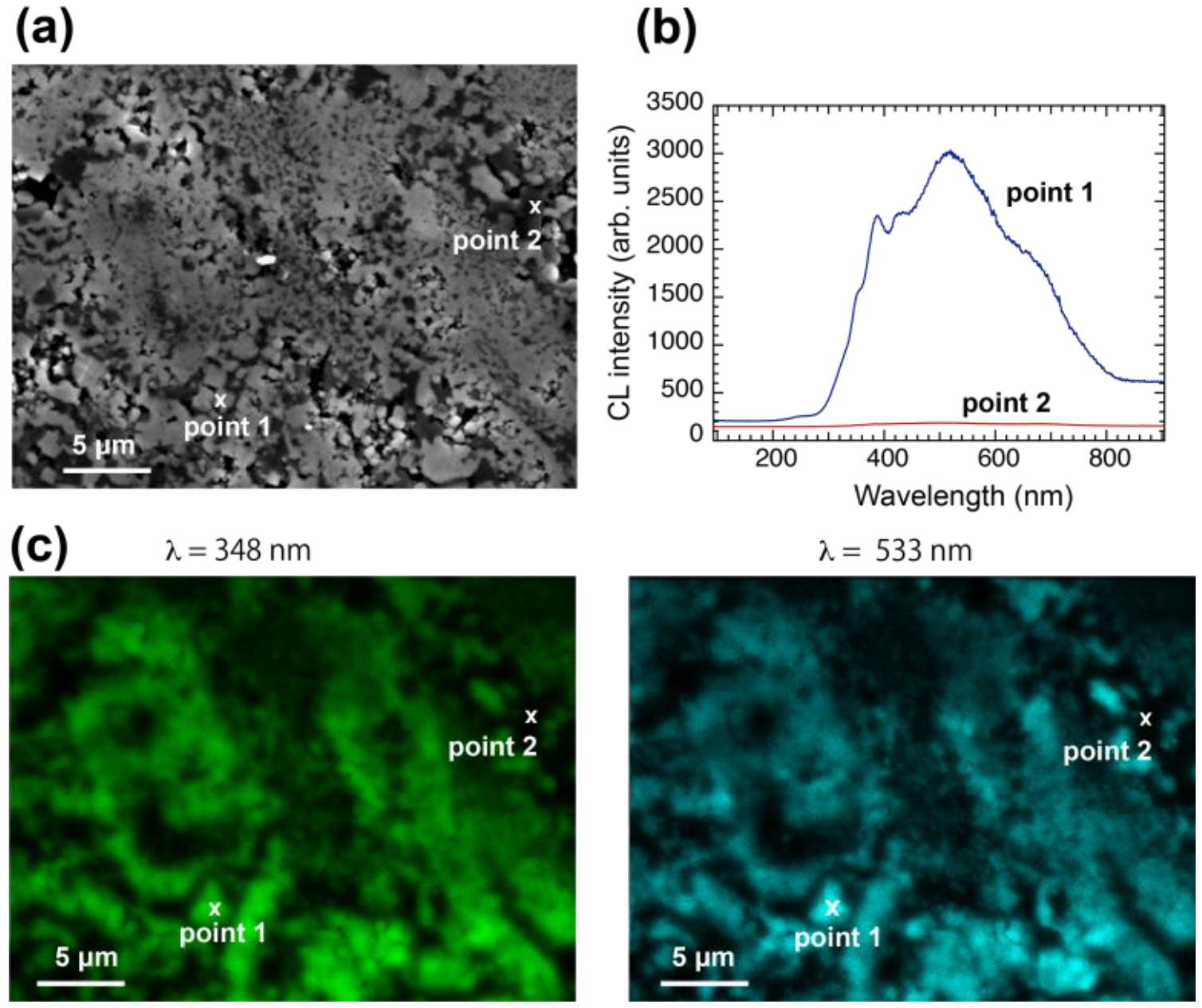

e
0
1

FIG. S2. Spatial distribution of oxygen vacancies. (a) Scanning electron microscope (SEM) image (4000×) of the $MgO/MgB_2$ nanocomposite. The gray and black regions correspond to the MgO and $MgB_2$ rich regions, respectively (see also Extended Data Fig. 1). (b) The cathodoluminescence (CL) spectra acquired at points 1 and 2 in (a). The observed CL bands in the wavelength range from 300 to 800 nm are attributed to the emissions from a variety of oxygen vacancies in MgO (for details, see [38]). (c) False-color CL images obtained at wavelengths of 349 nm (left panel) and 533 nm (right panel). Oxygen vacancies surround the $MgB_2$-rich regions, not only providing conduction channels for electron and hole currents but also contributing to the Andreev reflections at the MgO/ $MgB_2$ interfaces.

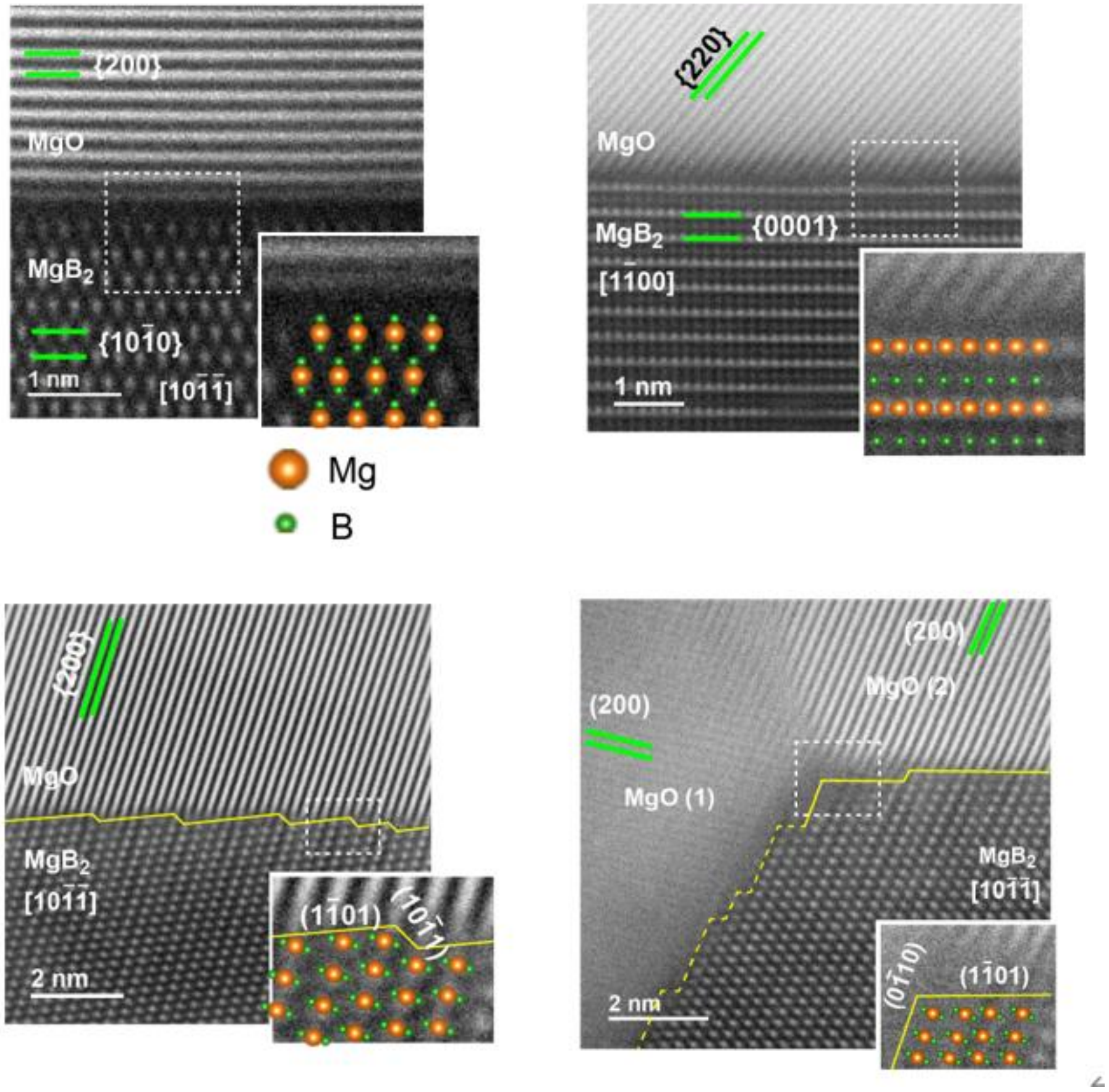

FIG. S3. High-angle annular dark field scanning transmission electron microscope (HAADF-STEM) images of various $MgO/MgB_2$ interfaces in the $MgO/MgB_2$ nanocomposite. The insets indicate the enlarged images of the corresponding white box regions, along with possible atomic arrangements of Mg (orange) and B (green) atoms. The figures are adopted from [38]. Copyright 2025 Author(s), licensed under a Creative Commons Attribution (CC BY) License.

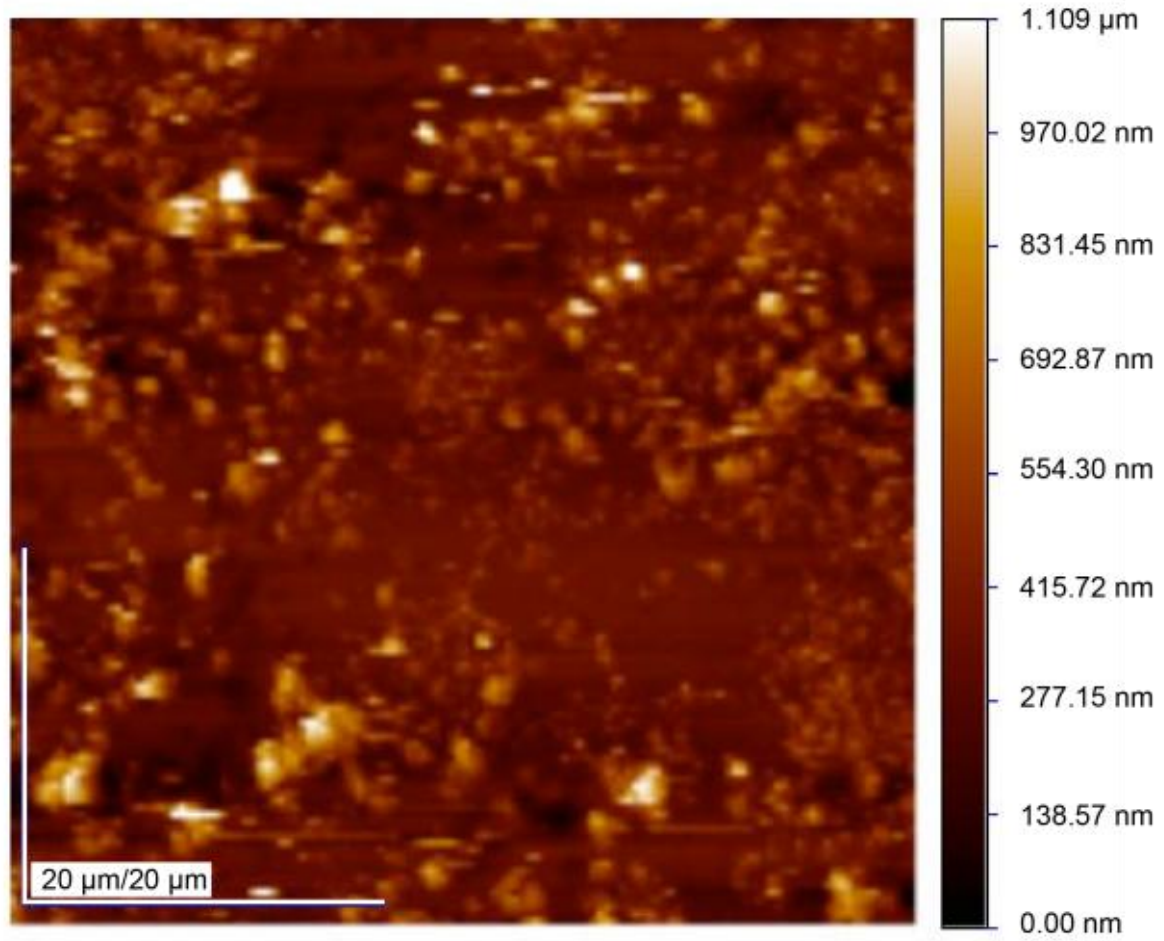

FIG. S4. The AFM characterization was carried out by JEOL JSPM-5400. The region to be measured was chosen randomly and scanned at ~4 µm/s scan rate. The image size is 50 × 50 µm with 128 × 128 pixels. For the AFM sample, a small amount of Mn (less than 0.1 wt %) was intentionally included for some practical reason.

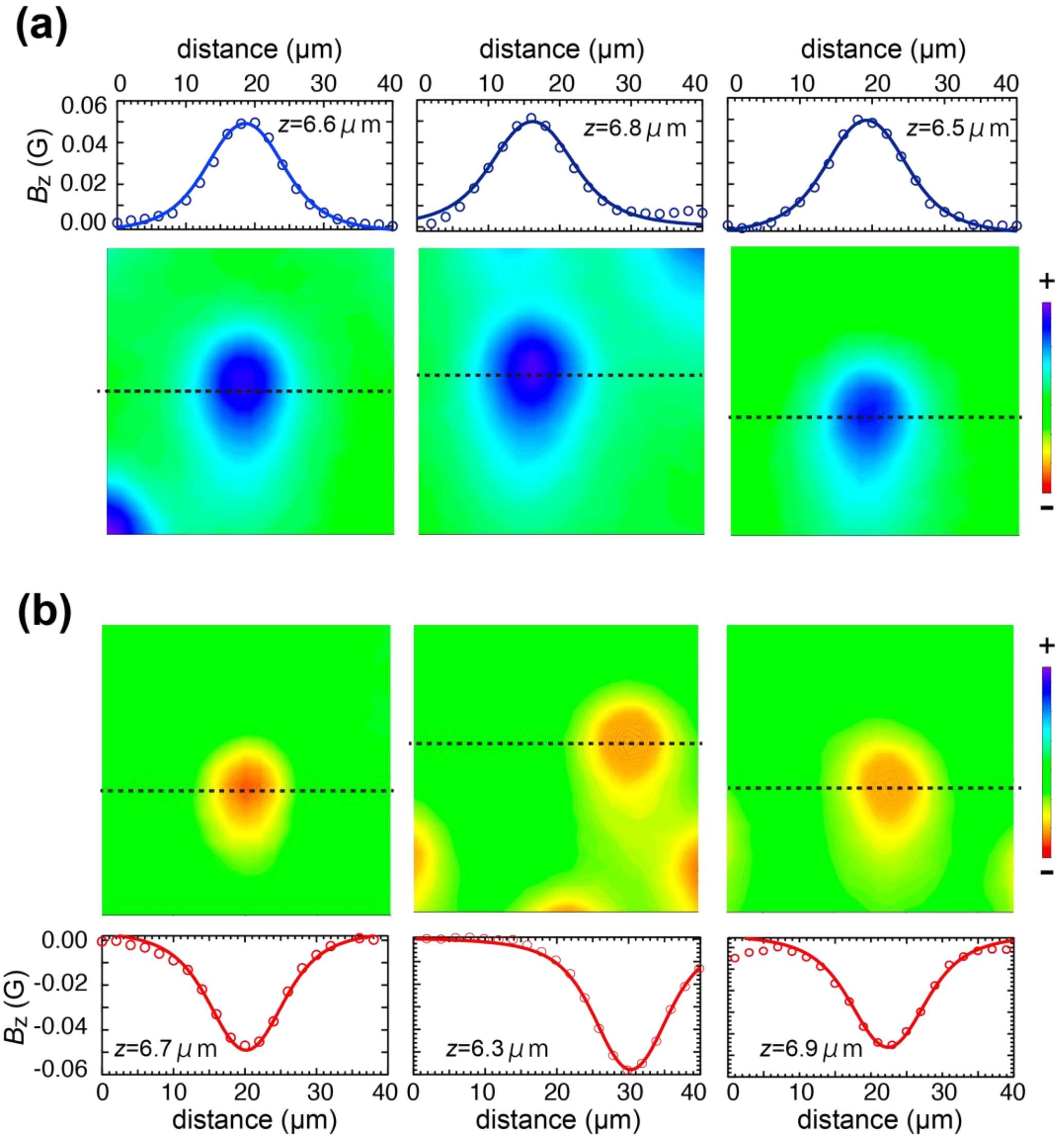

FIG. S5. Monopole model fitting for various isolated vortices. SSM images for isolated vortices obtained under (a) positive field cooling and (b) negative field cooling processes. The corresponding field profiles along a line scan indicated by a dotted line in the respective SSM images are also shown. Open circles represent the data, and the solid line is the fit to the monopole model for the flux quantum. The values of the distance from the sample surface to the pickup loop $z$ are obtained by assuming that $\lambda$ is negligibly smaller than $z$.